\documentclass[10pt,aps,prd,twocolumn,showpacs,amsmath,amssymb,nofootinbib,preprintnumbers,superscriptaddress]{revtex4-2}

\usepackage{xcolor}
\usepackage{mathtools}					
\usepackage[colorlinks=true,
            citecolor=red,
            linkcolor=blue,
            urlcolor=violet,
            filecolor=cyan,
            backref=false]{hyperref}
\usepackage{tikz}
\usetikzlibrary{calc,positioning,arrows.meta,decorations.pathmorphing,fit,backgrounds}
\tikzset{every picture/.style={line width=0.75pt}}

\newcommand{\lie}{\pounds}
\newcommand\bs{\boldsymbol}
\renewcommand{\cosh}{\operatorname{ch}}
\renewcommand{\sinh}{\operatorname{sh}}
\renewcommand{\tanh}{\operatorname{th}}
\renewcommand{\coth}{\operatorname{cth}}

\newcommand\feq{\mathrel{\phantom{=}}}

\newcommand{\xfrac}[2]{{\scriptstyle\frac{#1}{#2}}}

\begin{document}
\title{Double Wick rotations between symmetries of Taub--NUT, near-horizon~extreme~Kerr, and swirling spacetimes}

\author{Aimeric Coll\'{e}aux}
\email{aimeric.colleaux@matfyz.cuni.cz}
\affiliation{Institute of Theoretical Physics, Faculty of Mathematics and Physics, Charles University, V Hole\v{s}ovi\v{c}k\'ach 2, Prague 180 00, Czech Republic}

\author{Ivan Kol\'a\v{r}}
\email{ivan.kolar@matfyz.cuni.cz}
\affiliation{Institute of Theoretical Physics, Faculty of Mathematics and Physics, Charles University, V Hole\v{s}ovi\v{c}k\'ach 2, Prague 180 00, Czech Republic}

\author{Tom\'a\v{s} M\'alek}
\email{malek@math.cas.cz}
\affiliation{Institute of Mathematics of the Czech Academy of Sciences, \v{Z}itn\'a 25, 115 67 Prague 1, Czech Republic}

\date{\today}

\begin{abstract}
We explicitly show that certain 4-dimensional infinitesimal group actions with 3-dimensional orbits are related by double Wick rotations. In particular, starting with the symmetries of the spherical/hyperbolic/planar Taub--NUT spacetimes, one can obtain symmetries of the near-horizon extreme Kerr (NHEK) geometry or swirling universe by complex analytic continuations of coordinates. Similarly, the static spherical/hyperbolic/planar symmetries (i.e., symmetries of the Schwarzschild spacetime and other A-metrics) are mapped to symmetries of the B-metrics (or Melvin spacetime). All these mappings are theory-independent --- they constitute relations among symmetries themselves, and, hence among the classes of symmetry-invariant metrics and electromagnetic field strengths, rather than among specific solutions. Consequently, finding, e.g., vacuum Taub--NUT-type solutions in a given gravitational theory automatically yields vacuum NHEK- or swirling-type solutions of that theory, with a possible extension to the electromagnetic case.
\end{abstract}

\maketitle

\section{Introduction}

Several exact electro-vacuum solutions of general relativity (GR) are known to be related to each other via double Wick rotations in specific coordinates (sometimes also accompanied by complex analytic continuations of specific parameters). For example, it is straightforward to obtain the BI-metric (also known as the `bubble of nothing') from the Schwarzschild (Schw.) spacetime \cite{Witten:1981gj,Aharony:2002cx,Horowitz:2002cx}. Interestingly, the planar Reissner--Nordstr\"om (RN) spacetime can be mapped to the Melvin spacetime \cite{Astorino:2012zm}, while the massless Taub--NUT (TNUT) spacetime is related to the near-horizon extreme Kerr (NHEK) geometry \cite{Kunduri:2008tk}. Furthermore, the recently popular swirling universe can be obtained from the planar TNUT solution \cite{Astorino:2022aam} while the swirling--Melvin geometry is related to the charged planar TNUT~\cite{Barrientos:2024pkt}.

These double Wick rotations, however, were typically restricted to relations between particular solutions of specific theories. Although, the similarity between spacetimes was sometimes used as a guiding tool when searching for the corresponding solutions, e.g., in the nonlinear electrodynamics (NLE) when deriving Melvin-type solution from known RN-type solution \cite{Gibbons:2001sx} or the swirling-type solutions inspired by the known massless TNUT-type solution \cite{Barrientos:2024umq}, the direct theory-independent relations have never been analyzed in detail to the best of our knowledge.

In this paper, we aim to bridge this gap and demonstrate that in all the above cases (and beyond), it is the spacetime symmetries themselves (encoded in the infinitesimal group actions), rather than any specific solutions, that are related by double Wick rotations. Consequently, these relations work among entire classes of symmetry-invariant metrics (and possibly other symmetry-invariant tensor fields) and are completely independent of a given theory. They map a spacetime with a given symmetry to another spacetime with a different symmetry. If the former is a vacuum solution of a given metric theory of gravity, then so is the latter. In some cases, this correspondence extends to the electromagnetic sector (and possibly other dynamical fields), but it may require extra complex analytic continuation of parameters.

One reason this link has not been spelled out explicitly in the literature could be that the modern mathematical classification of infinitesimal group actions based on Lorentzian Lie algebra-subalgebra pairs \cite{Fels_Renner_2006,Bowers:2012,Snobl2014-te,Rozum2015-lp,Hicks:thesis} (see also \cite{Frausto:2024egp}) is somewhat outside the standard toolkit of most theoretical physicists. Particularly, the classification due to Hicks \cite{Hicks:thesis}, contains a comprehensive list of symmetry-invariant metrics that can be checked for similar relations. Here, we focus primarily on a physically attractive subset of [4,3,-] which admit 4-dimensional Lie algebras and 3-dimensional orbits.

The paper is structured as follows: We begin by writing the infinitesimal group actions [4,3,\{1--6\}] covering, e.g, the symmetries of spherical/hyperbolic/planar Schw. (also known as the A-metrics) \cite{EhlersKundt1962,Gott1974,Podolsky:2018dpr,Hruska:2018djo} and TNUT \cite{Taub:1950ez,Newman:1963yy,Griffiths:2009dfa} in a unified form, and provide the corresponding symmetry-invariant metrics and electromagnetic field strengths in Sec.~\ref{sc:TNUT}. The main part is the explicit complex analytic continuation of these infinitesimal group actions via double Wick rotations, along with the corresponding relations for symmetry-invariant tensors [4,3,\{8--11\}], in Sec.~\ref{sc:dw}. These infinitesimal group actions describe, for example, the symmetries of BI/BII/BIII-metrics \cite{EhlersKundt1962,Gott1974} (or Melvin spacetime \cite{Bonnor:1954tis,Melvin:1963qx}), NHEK \cite{Bardeen:1999px,Kunduri:2007vf}, swirling spacetime \cite{Gibbons:2013yq,Astorino:2022aam}. The obtained relations are then discussed on specific known examples in GR and NLE in Sec.~\ref{sc:application}. Finally, we present our conclusions in Sec.~\ref{sc:concl}. For the reader’s convenience, we include a brief review of the Hicks classification of infinitesimal group actions in App.~\ref{sc:Hicks}.

\paragraph*{Notation} We adopt the mostly-plus signature, i.e., $(-,+,+,+)$, for the Lorentzian metric $\bs{g}$. Abstract tensor indices are suppressed, and tensors are written in \textbf{boldface}. The symbol $\bs{\mathrm{d}}$ denotes the exterior derivative, and $\lie$ the Lie derivative. The symbols $\vee$ and $\wedge$ denote the the symmetric and antisymmetric (exterior) products with the conventions $\bs{\alpha}\vee\bs{\beta}=\bs{\alpha}\bs{\beta}+\bs{\beta}\bs{\alpha}$, $\bs{\alpha}\wedge\bs{\beta}=\bs{\alpha}\bs{\beta}-\bs{\beta}\bs{\alpha}$, for 1-forms $\bs{\alpha}$ and $\bs{\beta}$.

\section{[4,3,\{1--6\}] --- symmetries of Taub--NUT and Schwarzschild}\label{sc:TNUT}

Let us consider the infinitesimal group action $\Gamma$ given by the following vector fields:\footnote{These are related to vector fields $\bs{Y}_i$ from \cite{Frausto:2024egp} by means of:  { $\{\bs{X}_1,\bs{X}_2,\bs{X}_3,\bs{X}_4\} =
\big({[4,3,1]\!\!: \{\bs{Y}_1,\bs{Y}_2,-\bs{Y}_3,\bs{Y}_4\}},\\
{[4,3,2]\!\!: \{\bs{Y}_1,\bs{Y}_2,-\bs{Y}_3-\bs{Y}_4,\frac{\bs{Y}_4}{2n}\}},
{[4,3,3]\!\!: \{\bs{Y}_1,-\bs{Y}_2,\bs{Y}_3,\bs{Y}_4\}},\\
{[4,3,4]\!\!: \{\bs{Y}_1,-\bs{Y}_2,\bs{Y}_3+\bs{Y}_4,\frac{\bs{Y}_4}{2n}\}},
{[4,3,5]\!\!: \{\bs{Y}_3,-\bs{Y}_2,\bs{Y}_4,\frac{\bs{Y}_1}{2n}\}},\\
{[4,3,6]\!\!: \{\bs{Y}_1,-\bs{Y}_2,-\bs{Y}_3,\bs{Y}_4\}}\big)$} with the coordinate transformations:
${y_1 = \tfrac{t}{2n} {+} k \varphi}$,  ${y_2 = r}$, ${y_3 = \sin^{-1}\rho \text{ or } \sinh^{-1}\rho}$, ${y_4 = k \varphi}$ for ${k=\pm1}$, and the transformation ${y_1=\tfrac{t}{2n}{-}\tfrac{\rho^2}{4}\sin(2\varphi)}$, ${y_2=r}$, ${y_3=\rho\cos\varphi}$, ${y_4=-\rho\sin\varphi}$ for ${k=0}$.}
\begin{equation}\label{eq:iga}
\begin{aligned}
    \bs{X}_1 &= \sqrt{1{-}k\rho ^2} \left[\cos \varphi\bs{\partial}_\rho{-}\tfrac{  \sin \varphi}{\rho }\bs{\partial}_\varphi\right]{-}2n\tfrac{1{-}\sqrt{1{-}k\rho ^2} }{k}\tfrac{\sin \varphi}{\rho}\bs{\partial}_t\,, 
    \\ 
    \bs{X}_2 &=\sqrt{1{-}k\rho ^2}\left[ \sin \varphi\bs{\partial}_\rho{+}\tfrac{\cos \varphi }{\rho }\bs{\partial}_{\varphi}\right]{+}2n\tfrac{1{-}\sqrt{1{-}k\rho ^2} }{k}\tfrac{\cos \varphi}{\rho}\bs{\partial}_t\,, 
    \\
    \bs{X}_3 &=\bs{\partial}_{\varphi}\,, \quad \bs{X}_4 =\bs{\partial}_{t}\,,
\end{aligned}
\end{equation}
where ${k=\pm1,0}$, ${n\in\mathbb{R}}$, and ${(t,r,\rho,\varphi)}$ are some coordinates on the manifold. (At the level of infinitesimal group actions, all nonzero values of $n$ are equivalent; we keep $n$ explicitly only to allow a well-defined limit ${n \to 0}$.) The case ${k=0}$ is meant in the limiting sense ${k\to0}$, where ${(1-\sqrt{1-k\rho ^2} )/k\to\rho^2/2}$.  Using the terminology of the Hicks classification \cite{Hicks:thesis,Frausto:2024egp}, which is summarized in App.~\ref{sc:Hicks}, this infinitesimal group action with ${n\neq0}$ is denoted by [4,3,4] for ${k=1}$, [4,3,2] for ${k=-1}$, or [4,3,5] for ${k=0}$; it describes symmetries of spherical/hyperbolic/planar TNUT spacetime. The case ${n=0}$ reduces to the static spherical [4,3,3] (${k=1}$), hyperbolic [4,3,1] (${k=-1}$), or planar symmetry [4,3,6] (${k=0}$), i.e., the symmetries of spherical/hyperbolic/planar Schw. spacetimes also known as the AI/AII/AIII-metrics. (Note that various solutions of different theories exist with these symmetries; these GR solutions are mentioned only for illustration and to introduce convenient terminology, namely 'symmetries of ...'.) All of these infinitesimal group actions belong to a larger class [4,3,-], which admits exactly 4 independent Killing vector fields and have 3-dimensional orbits.

The most general $\Gamma$-invariant metric $\bs{g}$, ${\lie_{\bs{X}_i}{\bs{g}}=0}$, $\forall i$, i.e., the metric of which \eqref{eq:iga} is the Lie algebra of Killing vector fields, can be written in the form\footnote{When comparing the $\Gamma$-invariant metrics to \cite{Frausto:2024egp}, one has to perform the following redefinitions: ${\phi_1=(2n)^2ab}$, ${\phi_2= 2nd}$, ${\phi_3= \tfrac{1}{a}}$, ${\phi_4= c}$.}
\begin{equation}\label{Ansatz}
\begin{aligned}
    \bs{g} &= - a(r) b(r)\left(\bs{\mathrm{d}}t +2n \bs{\omega}_k \right)^2   + \frac{\bs{\mathrm{d}}r^2}{a(r)} + c(r) \bs{q}_{k}
    \\
    &\feq+d(r)\left(\bs{\mathrm{d}}t + 2n \bs{\omega}_{k} \right)\vee\bs{\mathrm{d}}r\;,
\end{aligned}
\end{equation}
where $\bs{q}_{k}$ is the 2-dimensional Euclidean metric of constant curvature, i.e., 2-sphere (${k=1}$), hyperbolic 2-space (${k=-1}$), or 2-plane (${k=0}$), 
\begin{equation}
        \bs{q}_{k} = \frac{\bs{\mathrm{d}}\rho^2}{1-k \rho^2} + \rho^2 \bs{\mathrm{d}}\varphi^2 \,,
    \quad
\end{equation}
and $\bs{\omega}_{k}$ is the 1-form given~by
\begin{equation}
    \bs{\omega}_{k} = \frac{1 - \sqrt{1-k \rho^2}}{k} \bs{\mathrm{d}}\varphi \,.
\end{equation}
The Lorentzian signature requires ${b+d^2>0}$ and ${c>0}$. Similarly, the most general $\Gamma$-invariant electromagnetic field strength tensor $\bs{F}$, ${\lie_{\bs{X}_i}{\bs{F}}=0}$, $\forall i$, is given by
\begin{equation}\label{Ansatz2}
\begin{aligned}
    \bs{F} &= \frac{\rho f_1(r)}{\sqrt{1 - k\rho^2}} \, \bs{\mathrm{d}}\rho \wedge \bs{\mathrm{d}}\varphi + f_2(r) \left( \bs{\mathrm{d}}t + 2n \bs{\omega}_k \right) \wedge \bs{\mathrm{d}}r\;, 
    \\
    f_1' &=-2nf_2\;,
\end{aligned}
\end{equation}
where the extra condition on the second line corresponds to $\bs{F}$ being closed, ${\bs{\mathrm{d}\bs{F}}=0}$.\footnote{Remark that the corresponding $\bs{A}$, $\bs{F}=\bs{\mathrm{d}}\bs{A}$, need not be $\Gamma$-invariant. For $n\neq0$, every closed $\Gamma$-invariant 2-form \eqref{Ansatz2} is completely determined by the exterior derivative of a $\Gamma$-invariant 1-form $\bs{A} = A_1(r)\,( \bs{\mathrm{d}}t + 2n \bs{\omega}_k) + A_2(r)\,\bs{\mathrm{d}}r$, with $f_1 = 2n A_1$ and $f_2 = - A_1'$, where the $A_2$ term is a pure gauge. In contrast, for $n=0$ the exterior derivative of a $\Gamma$-invariant 1-form $\bs{A} = A_1(r)\,\bs{\mathrm{d}}t + A_2(r)\,\bs{\mathrm{d}}r $ contributes only to the $f_2$ component of the 2-form \eqref{Ansatz2}, with $f_2 = -A_1'$. A general closed $\Gamma$-invariant 2-form can then be obtained only by including a non-invariant part in the potential 1-form, for example $\bar{\bs{A}} = \bs{A} + f_1 \bs{\omega}_k$, where $f_1$ is now a constant. The non-invariant part corresponds physically to magnetic monopoles.}

Let us stress that we will be primarily interested in the properties of the general $\Gamma$-invariant metrics \eqref{Ansatz} and $\Gamma$-invariant field strengths \eqref{Ansatz2} with arbitrary functions $a$, $b$, $c$, $d$, $f_1$, and $f_2$ rather than any specific solutions of some theory.

In general, it is possible to fix two of the four metric functions, e.g., ${c=r^2+n^2}$, ${d=0}$, (other useful alternative is ${b=1}$, ${d=0}$) due to the freedom in the form of the metric ansatz \eqref{Ansatz}, which is fully captured by the residual diffeomorphism subgroup. Its generators are the vector fields $\bs{\mathcal{W}}$ satisfying ${[\bs{\mathcal{W}},\bs{X}_i]=\sum_{j=1}^{4} a_{ij} \bs{X}_j}$, ${ i=1,\dots,4}$, for some constants $a_{ij}$. Then the flow $\Phi_{\tau}$ of $\bs{\mathcal{W}}$ transforms $\bs{g}$ to $\Phi_{\tau}^*\bs{g}$, so that it takes again the form of the $\Gamma$-invariant metric \eqref{Ansatz}.\footnote{This is because  $\lie_{\bs{\mathcal{W}}}\bs{g}$ is $\Gamma$-invariant, ${\lie_{\bs{X}_i}\lie_{\bs{\mathcal{W}}}\bs{g}=0}$, $\forall i$, as a consequence of $[\bs{\mathcal{W}},\bs{X}_i]$ being a Killing vector again.} For example, the Killing vectors $\bs{X}_i$ themselves generate trivial residual diffeomorphisms that do not change $a$, $b$, $c$, $d$. Those that induce non-trivial transformations of $a$, $b$, $c$, $d$, can be split into the $\Gamma$-invariant ones, ${[\bs{\mathcal{V}},\bs{X}_i]=0}$, $\forall i$, and non-$\Gamma$-invariant ones, ${[\bs{\mathcal{P}},\bs{X}_i]\neq0}$. Specifically, one finds
\begin{equation}\label{eq:resdifgen}
\begin{aligned}
    \bs{\mathcal{V}} &=\mathcal{V}^t(r)\, \bs{\partial}_{t} \allowbreak + \mathcal{V}^r(r)\, \bs{\partial}_{r}\;,
\\
    \bs{\mathcal{P}} &=\begin{cases}
        t\bs{\partial}_{t}\;, \rho\bs{\partial}_{\rho}\;, &n=0\;, \; k=0\;,
        \\
        t\bs{\partial}_{t}\;, &n=0\;, \; k=\pm1\;,
        \\
        t\bs{\partial}_{t}+\tfrac{\rho}{2}\bs{\partial}_{\rho} \;, &n\neq 0\;, \; k=0\;,
        \\
        \emptyset \;, & n\neq 0\;, \; k=\pm1\;,
    \end{cases}
\end{aligned}
\end{equation}
where the two arbitrary functions $\mathcal{V}^t(r)$ and $\mathcal{V}^r(r)$ in the generator $\bs{\mathcal{V}}$ describe exactly the possibility of the above gauge-fixings, while the generators $\bs{\mathcal{P}}$ give rise to a constant scaling.\footnote{In the context of symmetry reduction of gravitational Lagrangians, the existence of generators $\bs{\mathcal{P}}$ together with Noether identities associated to generators $\bs{\mathcal{V}}$, may, in some cases, justify the possibly problematic gauge fixing at the level of the reduced Lagrangians \cite{Frausto:2024egp}.}

There is one remark to be made about the coordinates used above in the case ${k=1}$. The metric with ${\rho\in(0,1)}$ only covers the half of the spacetime with the regular semi-axis located at ${\rho=0}$. The other half of the spacetime can be obtained by formal replacement ${\bs{\omega}_{1} \;\to\; \big(1 {+} \sqrt{1{-}\rho^2}\big) \bs{\mathrm{d}}\varphi}$, where again ${\rho\in(0,1)}$, but ${\rho=0}$ now covers the other semi-axis with the Misner string. This can be seen as follows: For any $k$, we assume that the coordinate $\varphi$ is $2\pi$ periodic. This is equivalent to ${\bs{Y}=\bs{\partial}_{\varphi}}$ being a Killing vector with closed orbits whose flow parameter has the period ${2\pi}$. If ${k=1}$, then the region containing closed timelike curves satisfies ${\bs{Y}^2=g_{\varphi\varphi}=-(2n)^2a(r)b(r)\big({1 \mp \sqrt{1{-}\rho^2}}\big)^2+c(r)\rho^2<0}$, where $\mp$ stands for original/new $\bs{\omega}_{1}$, respectively. The fact that this is satisfied for the $+$ sign near ${\rho=0}$ where ${a(r)>0}$ (assuming ${d=0}$, which implies ${b>0}$ for the Lorentzian signature) signifies the presence of the Misner string.

\section{Complex analytic continuation of symmetries}\label{sc:dw}
In what follows, we show that the infinitesimal group actions [4,3,\{1--6\}] given by \eqref{eq:iga} can be directly mapped, via double Wick rotations, to the infinitesimal group actions [4,3,\{8--10\}] describing the symmetries of the BI/BII/BIII-metrics (or Melvin spacetime), the NHEK geometry, or the swirling spacetime. This leads to theory-independent relations within the entire classes of $\Gamma$-invariant metrics and $\Gamma$-invariant field strengths. A schematic illustration of these mappings via double Wick rotations is shown in Fig.~\ref{fig:enter-label}.

\begin{figure*}[ht]
    \centering
    \begin{tikzpicture}[node distance=7mm and 17mm,
        shorten >=1mm, shorten <=1mm,
        box/.style={rectangle,
            minimum width=31mm,
            minimum height=12mm,
            rounded corners=1mm,
            thick,
            draw,
            align=center
        },
        wick/.style={{Stealth[length=0pt 4 0, fill=white]}-{Stealth[length=0pt 4 0, fill=white]}, draw, thin,
            shorten <=1pt, shorten >=1pt,
            double, double distance=1pt,
            decorate, decoration={snake, amplitude=0.2mm, segment length=1mm, post length=1.5mm, pre length=2mm}
            }
        ]
        
        \node (438) [box]   {[4,3,8]\\\scriptsize \textsf{ch.\,BI/II--(A)dS}};
        \node (433) [box,right=of 438, label=below:{$k=1$}]  {[4,3,3]\\ {\scriptsize \textsf{sph.\,RN--(A)dS}}};
        \node (431) [box,below=of 433, label=below:{$k=-1$}]  {[4,3,1]\\ {\scriptsize \textsf{hyp.\,RN--(A)dS}}};
        \node (436) [box,below=of 431, label=below:{$k=0$}]  {[4,3,6]\\ {\scriptsize \textsf{pl.\,RN--(A)dS}}};
        \node (4311) [box,left=of 436]  {[4,3,11]\\{\scriptsize \textsf{Melvin--(A)dS}}};
        \node (434) [box,right=of 433, label=below:{$k=1$}]  {[4,3,4]\\ {\scriptsize \textsf{ch.\,sph.\,TNUT--(A)dS}}};
        \node (432) [box,right=of 431, label=below:{$k=-1$}]  {[4,3,2]\\ {\scriptsize \textsf{ch.\,hyp.\,TNUT--(A)dS}}};
        \node (435) [box,right=of 436, label=below:{$k=0$}]  {[4,3,5]\\ {\scriptsize \textsf{ch.\,pl.\,TNUT--(A)dS}}};
        \node (439) [box,right=of 434]  {[4,3,9] \\{\scriptsize \textsf{hht./sht. NHEK--N--(A)dS}}};
        \node (4310) [box,right=of 435] {[4,3,10]\\{\scriptsize \textsf{swirling--Melvin--(A)dS}}};

        \path[-Stealth] (434) edge node[above] {} (433)
                        (432) edge node[above] {} (431)
                        (435) edge node[above] {} (436);

        \draw [-Stealth, rounded corners=1mm] (439) -- ++(0,1.7) -- node[above] {} ($(438)+(0,1.7)$) -- (438);
        
        \draw [-Stealth, rounded corners=1mm ] (4310) -- ++(0,-1.7) -- node[above] {} ($(4311)-(0,1.7)$) -- (4311);

        \path[<->, every edge/.style={wick}]
            (438) edge node[pos=0.5, above] {$+$} (433)
            (438) edge node[pos=0.5, above] {$-$} (431)
            (436) edge (4311)
            (434) edge node[pos=0.5, above] {$+$} (439)
            (432) edge node[pos=0.5, above] {$-$} (439)
            (435) edge (4310);

        \begin{scope}[on background layer]
            \node[draw, thin, densely dashed, fit=(434) (432) (435),
      inner sep=6mm, rounded corners=1mm,
      label=below:{$n\neq0$}] {};
            \node [draw, thin, densely dashed, fit=(433) (431) (436), inner sep=6mm, rounded corners=1mm, label=below:{$n=0$}] {};
        \end{scope}
    \end{tikzpicture}
    \caption{Diagram of possible double Wick rotations, indicated by double wiggly arrows (\!\protect\tikz[baseline=-0.5ex]{\protect\draw[{Stealth[length=0pt 4 0, fill=white]}-{Stealth[length=0pt 4 0, fill=white]}, draw, thin, shorten <=1pt, shorten >=1pt, double, double distance=1pt, decorate, decoration={snake, amplitude=0.2mm, segment length=1mm, post length=1.5mm, pre length=2mm}] (0,0)--(2.5em,0);}\!), among infinitesimal group actions denoted by ${[d,l,c]}$ according to the Hicks classification \cite{Hicks:thesis,Frausto:2024egp} (summarized in App.~\ref{sc:Hicks}) and the corresponding classes of $\Gamma$-invariant tensors (metrics and field strengths). The signs $+$ and $-$ above these arrows indicate that the maps of the $\Gamma$-invariant metrics are restricted to $\mathrm{dS_2}$ and $\mathrm{AdS_2}$ structures (in metrics of [4,3,\{8,9\}]), respectively. The simple arrows (\protect\tikz[baseline=-0.5ex]{\protect\draw[-Stealth,thin] (0,0)--(1.8em,0);}) correspond to the limits from the actions with ${n\neq0}$ (equivalent for different non-zero values of $n$) to ${n=0}$. Description in \textsf{sans-serif} provides examples of spacetimes with such symmetries within electro-vacuum solutions of Einstein--Maxwell--$\Lambda$ theory. The shorthands stand for spherical (sph.), hyperbolic (hyp.), planar (pl.), charged (ch.), spherical horizon topology (sht.), and hyperbolic horizon topology (hht.).}
    \label{fig:enter-label}
\end{figure*}

\subsection{[4,3,8] --- symmetries of BI/BII}

First we consider the case ${n=0}$. The infinitesimal group action [4,3,8] is obtained from ${k=1}$ case (i.e., [4,3,3]) by considering ${\rho>1}$ instead of ${|\rho|<1}$ (this can be thought of as a Wick rotation: ${\vartheta= i\theta}$ for ${\rho=\sin \vartheta}$) together with the Wick rotation ${t=iq}$. The real vector fields ${-i\bs{X}_1}$, ${i\bs{X}_2}$, ${\bs{X}_3}$,  ${i\bs{X}_4}$, then give rise to the infinitesimal group action [4,3,8].\footnote{$\{\mp i\bs{X}_1, \pm i\bs{X}_2, \bs{X}_3, i\bs{X}_4\} = \{\bs{Y}_1,\bs{Y}_2,\bs{Y}_3,\bs{Y}_4\}$ for $y_2>0$ and $y_2<0$, respectively, under the coordinate transformation $q=y_4$, $r=y_3$, $\rho=\cosh y_2$, $\varphi=y_1$ \cite{Frausto:2024egp}.} These correspond to the symmetries of, for example, the BI/BII-metrics.

Under the change of range of $\rho$, the Riemannian metric of the 2-sphere $\bs{q}_{+}$ becomes the Lorentzian metric of $\mathrm{dS_2}$, which we denote by $\check{\bs{q}}_{+}$,
\begin{equation}
    \check{\bs{q}}_{+}= -\frac{\bs{\mathrm{d}}\rho^2}{\rho^2-1} + \rho^2 \bs{\mathrm{d}}\varphi^2\,.
\end{equation}
Notice that $-\check{\bs{q}}_{+}$ corresponds to $\mathrm{AdS_2}$. Considering that the $\Gamma$-invariant metric of a real infinitesimal group action must be real, we are led to perform the redefinition ${\tilde{d}(r) = id(r)}$ so that the new function $\tilde{d}$ is real. In what follows, a tilde will denote such $i$-redefinitions of functions in this context. Hence, upon the double Wick rotation followed by $i$-redefinitions, the $\Gamma$-invariant metric reads
\begin{equation}\label{eq:metric438fromk433}
    \bs{g}_{\text{[4,3,8]}_+} = c(r) \check{\bs{q}}_{+}  + a(r) b(r) \bs{\mathrm{d}}q^2+ \frac{\bs{\mathrm{d}}r^2}{a(r)}  + \tilde{d}(r) \bs{\mathrm{d}}q \vee \bs{\mathrm{d}}r
\end{equation}
while the $\Gamma$-invariant field strength is
\begin{equation}\label{eq:F438fromk433}
    \bs{F}_{\text{[4,3,8]}_+}= -\frac{\rho \tilde{f}_1}{\sqrt{\rho^2-1}} \, \bs{\mathrm{d}}\rho \wedge \bs{\mathrm{d}}\varphi + \tilde{f}_2(r)  \bs{\mathrm{d}}q \wedge \bs{\mathrm{d}}r\;,
\end{equation}
where we similarly redefined ${\tilde{f}_1=if_1}$ and ${\tilde{f}_2(r)=if_2(r)}$. From the linearity of ${\lie_{\bs{X}}\bs{g}=0}$ and ${\lie_{\bs{X}}\bs{F}=0}$, it is not surprising that these are directly the general $\Gamma$-invariant tensors of the infinitesimal group action [4,3,8]. 

Alternatively, the same infinitesimal group actions can be also obtained from ${k=-1}$ case (i.e., [4,3,1]). First, we notice that $\bs{q}_-$ can be transformed to
\begin{equation}\label{eq:qminus}
    \bs{q}_-= \frac{\bs{\mathrm{d}}p^2}{p^2-1} + p^2 \bs{\mathrm{d}}h^2
\end{equation}
by means of the transformation ${\rho = \sqrt{\tanh ^2 h{+}p^2{-}1}\cosh h}$, ${\varphi= \cot ^{-1}\left({(p \sinh h)/\sqrt{p^2-1} }\right)}$, where ${p>1}$. Now, upon the Wick rotations ${h=i\psi}$, ${t=iq}$, the real vector fields ${i\bs{X}_1}$, ${\bs{X}_2}$, ${i\bs{X}_3}$, ${i\bs{X}_4}$ give rise to the infinitesimal group action [4,3,8].\footnote{${\{-i\bs{X}_1, -\bs{X}_2, i\bs{X}_3 , i\bs{X}_4 \} = \{\bs{Y}_3, \bs{Y}_1, \bs{Y}_2, \bs{Y}_4 \}}$ after the coordinate change $q=y_4$, $r=y_3$, $p=\cosh y_2$, $\psi=y_1$ \cite{Frausto:2024egp}.} As before, let us denote the resulting Lorentzian metric of $\mathrm{AdS_2}$ obtained from the Riemannian metric of the hyperbolic 2-space $\bs{q}_{-}$ by~$\check{\bs{q}}_{-}$,
\begin{equation}
    \check{\bs{q}}_{-}= - p^2 \bs{\mathrm{d}}\psi^2 +\frac{\bs{\mathrm{d}}p^2}{p^2-1} \,;
\end{equation}
again $-\check{\bs{q}}_{-}$ corresponds to $\mathrm{dS_2}$. Hence, the $\Gamma$-invariant metric is
\begin{equation}\label{eq:metric438from431}
    \bs{g}_{\text{[4,3,8]}_-} = c(r) \check{\bs{q}}_{-}  + a(r) b(r) \bs{\mathrm{d}}q^2+ \frac{\bs{\mathrm{d}}r^2}{a(r)}  + \tilde{d}(r) \bs{\mathrm{d}}q \vee \bs{\mathrm{d}}r
\end{equation}
and the $\Gamma$-invariant field strength becomes
\begin{equation}\label{eq:F438from431}
    \bs{F}_{\text{[4,3,8]}_-}= -\frac{p \tilde{f}_1}{\sqrt{p^2-1}} \, \bs{\mathrm{d}}p \wedge \bs{\mathrm{d}}\psi + \tilde{f}_2(r)  \bs{\mathrm{d}}q \wedge \bs{\mathrm{d}}r\;.
\end{equation}

Since ${\check{\bs{q}}_{+}=-\check{\bs{q}}_{-}}$ upon identifying the coordinates ${\rho=p}$, ${\varphi=\psi}$, the metrics \eqref{eq:metric438fromk433} and \eqref{eq:metric438from431} differ only by a formal replacement ${c(r)\to-c(r)}$. Although this is just a matter of redefinition at the level of the entire class of $\Gamma$-invariant metrics, it can become significant for the double Wick rotations of individual metrics, as we will discuss in Sec.~\ref{sc:application}.

It is easily seen that the BI/BII-metrics belong to both metric ansatzes $\bs{g}_{\text{[4,3,8]}_{\pm}}$ [(82), (87) from \cite{Podolsky:2018dpr}; using ${\rho=p=\sqrt{1+\hat{q}^2}}$].

\subsection{[4,3,11] --- symmetries of BIII (or Melvin)}

The infinitesimal group action [4,3,11], can be obtained from ${k=0}$ case (i.e., [4,3,6]) by ${t=iq}$, ${\rho=\sqrt{x^2-\tau^2}}$, ${\varphi=i\tanh ^{-1}\left({\tau }/{x}\right)}$. The real vector fields ${\bs{X}_1}$, ${i\bs{X}_2}$, ${i\bs{X}_3}$, ${i\bs{X}_4}$, then correspond to the infinitesimal group action [4,3,11].\footnote{$\{\pm\bs{X}_1, \pm i\bs{X}_2, i\bs{X}_3 , i\bs{X}_4 \} = \{\bs{Y}_2, \bs{Y}_1, \bs{Y}_4, \bs{Y}_3 \}$ for $x>0$ and $x<0$, respectively, upon relabeling the coordinates $q=y_4$, $r=y_3$, $x=y_2$, $\tau=y_1$ \cite{Frausto:2024egp}.} It captures the symmetries of, e.g., the BIII-metric or the Melvin universe. The Riemannian metric of the flat 2-space ${\bs{q}}_0$ becomes the Lorentzian metric of~$\mathbb{M}^2$,
\begin{equation}
    \check{\bs{q}}_0=-\bs{\mathrm{d}}\tau^2+\bs{\mathrm{d}}x^2\;.
\end{equation}
The $\Gamma$-invariant metric reads
\begin{equation}\label{eq:4311from436}
    \bs{g}_{\text{[4,3,11]}} = c(r) \check{\bs{q}}_{0} + a(r) b(r) \bs{\mathrm{d}}q^2+ \frac{\bs{\mathrm{d}}r^2}{a(r)}  + \tilde{d}(r) \bs{\mathrm{d}}q \vee \bs{\mathrm{d}}r
\end{equation}
and the $\Gamma$-invariant field strength is
\begin{equation}\label{eq:F4311from436}
    \bs{F}_{\text{[4,3,11]}}= -\tilde{f}_1 \, \bs{\mathrm{d}}\tau \wedge \bs{\mathrm{d}}x + \tilde{f}_2(r)  \bs{\mathrm{d}}q \wedge \bs{\mathrm{d}}r\;.
\end{equation}

The BIII metric is clearly included in the metric ansatz $\bs{g}_{\text{[4,3,11]}}$  [(90) from \cite{Podolsky:2018dpr}]. The same is true for the Melvin universe, as one can see by recasting \eqref{eq:4311from436}, via the coordinate transformation ${r=1+\hat{B}^2\hat{\rho}^2/4}$, ${\tau=\hat{t}}$, ${x=\hat{z}}$, ${q=2\hat{\phi}/\hat{B}^2}$ with some constant ${\hat{B}>0}$, to a more familiar form [(7.21) from \cite{Griffiths:2009dfa}],
\begin{equation}\label{eq:4311alternative}
\begin{aligned}
    \bs{g}_{\text{[4,3,11]}} &= c\big(1+\xfrac{\hat{B}^2\hat{\rho}^2}{4}\big) \big(-\bs{\mathrm{d}}\hat{t}^2+\bs{\mathrm{d}}\hat{z}^2\big) +\tfrac{\hat{B}^4\hat{\rho}^4\bs{\mathrm{d}}\hat{\rho}^2}{4a\big(1+\frac{\hat{B}^2\hat{\rho}^2}{4}\big)}
    \\
    &\feq+ \tfrac{4a\big(1+\xfrac{\hat{B}^2\hat{\rho}^2}{4}\big) b\big(1+\xfrac{\hat{B}^2\hat{\rho}^2}{4}\big)}{\hat{B}^4} \bs{\mathrm{d}}\hat{\phi}^2 
    \\
    &\feq+ \hat{\rho}\tilde{d}\big(1+\xfrac{\hat{B}^2\hat{\rho}^2}{4}\big) \bs{\mathrm{d}}\hat{\phi} \vee \bs{\mathrm{d}}\hat{\rho}\;.
\end{aligned}
\end{equation}

\subsection{[4,3,9] --- symmetries of NHEK}

Let us now move on to the ${n\neq 0}$ case. The infinitesimal group action [4,3,9] is obtainable from ${k=1}$ case (i.e., [4,3,4]) by taking ${\rho>1}$ and performing the complex transformation ${t=iq-2n\varphi}$. The real vector fields ${i\bs{X}_1}$, ${i\bs{X}_2}$, ${\bs{X}_3 - 2n \bs{X}_4}$, ${i\bs{X}_4}$ will correspond to the infinitesimal group action [4,3,9].\footnote{$\{i\bs{X}_1, -i\bs{X}_2, \bs{X}_3 - 2n \bs{X}_4, -2ni\bs{X}_4\} = \{\bs{Y}_1, \bs{Y}_2, \bs{Y}_3, \bs{Y}_4 \}$ for $y_2 > 0$ under the coordinate transformation $q=-2ny_3$, $r=y_4$, $\rho=\cosh y_2$, $\varphi=y_1$ \cite{Frausto:2024egp}.} These are the symmetries of, e.g., NHEK. The corresponding $\Gamma$-invariant metric then reads
\begin{equation}\label{eq:439from434}
\begin{aligned}
    \bs{g}_{\text{[4,3,9]}_+} &= c(r) \check{\bs{q}}_{+} + a(r) b(r)\left(\bs{\mathrm{d}}q {-}2n\sqrt{\rho^2{-}1}\bs{\mathrm{d}}\varphi \right)^2 + \frac{\bs{\mathrm{d}}r^2}{a(r)} \\
    &\quad  +\tilde{d}(r)\left(\bs{\mathrm{d}}q {-}2n\sqrt{\rho^2{-}1}\bs{\mathrm{d}}\varphi\right)\vee\bs{\mathrm{d}}r
\end{aligned}
\end{equation}
while the $\Gamma$-invariant field strength is
\begin{equation}\label{eq:F439from434}
\begin{aligned}
    \bs{F}_{\text{[4,3,9]}_+} &= -\frac{\rho \tilde{f}_1(r)}{\sqrt{\rho^2-1}} \, \bs{\mathrm{d}}\rho \wedge \bs{\mathrm{d}}\varphi \\
    &\quad -\frac{\tilde{f}'_1(r)}{2n}   \left( \bs{\mathrm{d}}q - 2n \sqrt{\rho^2-1}\bs{\mathrm{d}}\varphi \right) \wedge \bs{\mathrm{d}}r\;.
\end{aligned}
\end{equation}

Similar to the case [4,3,8], an alternative approach exists also for [4,3,9]. Indeed, the same infinitesimal group action [4,3,9] can be also obtained from ${k=-1}$ case (i.e., [4,3,2]) by means of the transformation ${\rho = \sqrt{\tanh ^2 h+p^2-1}\cosh h}$, ${\varphi= \cot ^{-1}\left({(p \sinh h)/\sqrt{p^2-1} }\right)}$ [as above to reach the form \eqref{eq:qminus}], but this time it must be accompanied by the transformation $t=s +2 n \cot ^{-1}\left((p \sinh h)/\sqrt{p^2-1}\right)-2 n \tan ^{-1}\left(\sqrt{p^2-1} \coth h\right)$, followed by Wick rotations ${s=iq}$, ${h=i\psi}$. Then the real vector fields ${i\bs{X}_1}$, ${\bs{X}_2}$, ${i(\bs{X}_3 + 2n \bs{X}_4})$, ${i\bs{X}_4}$ will again correspond to [4,3,9].\footnote{$\{-i\bs{X}_1, -\bs{X}_2, i(\bs{X}_3 + 2n \bs{X}_4), -2ni\bs{X}_4\} = \{\bs{Y}_3, \bs{Y}_1, \bs{Y}_2, \bs{Y}_4 \}$ for $y_2 > 0$ after the coordinate change $q=-2ny_3$, $r=y_4$, $p=\cosh y_2$, $\psi = y_1$ \cite{Frausto:2024egp}.} The $\Gamma$-invariant metric is then given by
\begin{equation}\label{eq:439from432}
\begin{aligned}
    \bs{g}_{\text{[4,3,9]}_-} &= c(r) \check{\bs{q}}_{-} + a(r) b(r)\left(\bs{\mathrm{d}}q {-}2n\sqrt{p^2{-}1}\bs{\mathrm{d}}\psi \right)^2 + \frac{\bs{\mathrm{d}}r^2}{a(r)} \\
    &\feq + \tilde{d}(r)\left(\bs{\mathrm{d}}q -2n\sqrt{p^2-1}\bs{\mathrm{d}}\psi\right)\vee\bs{\mathrm{d}}r
\end{aligned}
\end{equation}
and the $\Gamma$-invariant field strength by
\begin{equation}\label{eq:F439from432}
\begin{aligned}
    \bs{F}_{\text{[4,3,9]}_-} &= -\frac{p \tilde{f}_1(r)}{\sqrt{p^2-1}} \, \bs{\mathrm{d}}p \wedge \bs{\mathrm{d}}\psi \\
    &\quad -\frac{\tilde{f}'_1(r)}{2n}  \left( \bs{\mathrm{d}}q - 2n \sqrt{p^2-1}\bs{\mathrm{d}}\psi \right) \wedge \bs{\mathrm{d}}r\;.
\end{aligned}
\end{equation}
Again, upon identifying the coordinates ${\rho=p}$, ${\varphi=\psi}$, the metrics \eqref{eq:439from434} and \eqref{eq:439from432} are same up to a formal replacement ${c(r)\to-c(r)}$.

Despite being very natural, the above coordinates are rather non-standard. We can employ the transformation of coordinates 
${\hat{v}= 2n^2{(p \sin \psi+1)}/({\sqrt{p^2-1}-p \cos \psi})}$, ${\hat{r}= p \cos \psi-\sqrt{p^2-1}}$, $\hat{\phi} = {-}\tfrac{q}{2n}{-}\log \big(p \cos \psi{-}\sqrt{p^2{-}1}\big)\allowbreak {-}2 \tanh ^{-1}\big(\sqrt{\frac{p-1}{p+1}} \cot \psi{-}\csc \psi\big) {-}2 \tanh ^{-1}\tan \tfrac{\psi }{2}$, ${\cos\hat{\theta}= r/n}$, to recast \eqref{eq:439from432} to a more familiar form,
\begin{equation}\label{eq:439alt}
\begin{aligned}
    \bs{g}_{\text{[4,3,9]}_-} &=\tfrac{c\big(\frac{\hat{r}_0\cos\hat{\theta}}{\sqrt{2}}\big)}{\hat{r}_0^2}\big(-\tfrac{\hat{r}^2}{\hat{r}_0^2}\bs{\mathrm{d}}\hat{v}^2+\bs{\mathrm{d}}\hat{v}\vee\bs{\mathrm{d}}\hat{r}\big)+\tfrac{\hat{r}_0^2 \sin^2\hat{\theta}\,\bs{\mathrm{d}}\hat{\theta}^2}{2a\big(\frac{\hat{r}_0\cos\hat{\theta}}{\sqrt{2}}\big)}
    \\
    &\feq+2 \hat{r}_0^2 a\big(\xfrac{\hat{r}_0\cos\hat{\theta}}{\sqrt{2}}\big)b\big(\xfrac{\hat{r}_0\cos\hat{\theta}}{\sqrt{2}}\big)\big(\bs{\mathrm{d}}\phi+\tfrac{\hat{r}}{\hat{r}_0^2}\bs{\mathrm{d}}\hat{v}\big)^2
    \\
    &\feq+\hat{r}_0^2\tilde{d}\big(\xfrac{\hat{r}_0\cos\hat{\theta}}{\sqrt{2}}\big)\sin\hat{\theta}\big(\bs{\mathrm{d}}\hat{\phi}+\tfrac{\hat{r}}{\hat{r}_0^2}\bs{\mathrm{d}}\hat{v}\big)\vee\bs{\mathrm{d}}\hat{\theta}\;,
\end{aligned}
\end{equation}
where we denoted ${\hat{r}_0=\sqrt{2}n}$. Clearly, the NHEK belongs to both metric ansatzes $\bs{g}_{\text{[4,3,9]}_{\pm}}$ [(1) from \cite{Kunduri:2007vf}]. However, we expect this class to also contain the near-horizon extreme limits of the hyperbolic rotating black holes from \cite{Klemm:1997ea}, which do not seem to appear explicitly in the literature. However, their physicality is at least partially excluded by horizon topology theorems \cite{Kunduri:2008rs}.

\subsection{[4,3,10] --- symmetries of swirling}

The infinitesimal group action [4,3,10] can be obtained from ${k=0}$ case (i.e., [4,3,5]) by the Wick rotation ${t=i(q +nx\tau)}$, ${\rho=\sqrt{x^2-\tau^2}}$, ${\varphi=i\tanh ^{-1}\left({\tau }/{x}\right)}$. Then the real vector fields ${\bs{X}_1}$, ${i\bs{X}_2}$, ${i\bs{X}_3}$, ${i\bs{X}_4}$, will correspond to the infinitesimal group action [4,3,10].\footnote{$\{i\bs{X}_4, \pm \frac{\bs{X}_1 + i\bs{X}_2}{\sqrt{2}}, \pm \frac{i\bs{X}_2 - \bs{X}_1}{2n\sqrt{2}}, -i\bs{X}_3\} = \{\bs{Y}_1, \bs{Y}_2, \bs{Y}_3, \bs{Y}_4 \}$ for $x > 0$ and $x<0$, respectively, upon performing the coordinate transformation $q=y_1 - \frac{n}{4}(y_4 + \frac{1}{2n}y_3)(y_4 + \frac{1}{2n}y_3 + 2(y_4 -\frac{1}{2n}y_3)) + \frac{n}{4}(y_4 - \frac{1}{2n}y_3)^2$, $r=y_2$, $x=\frac{1}{\sqrt{2}}(y_4 - \frac{1}{2n}y_3)$, $\tau=\frac{1}{\sqrt{2}}(y_4 + \frac{1}{2n}y_3)$ \cite{Frausto:2024egp}.} It captures the symmetries of, for instance, the swirling universe. The corresponding $\Gamma$-invariant metric reads
\begin{equation}\label{eq:4310from435}
\begin{aligned}
    \bs{g}_{\text{[4,3,10]}} &= c(r) \check{\bs{q}}_{0} +  a(r) b(r)\left(\bs{\mathrm{d}}q +2nx\bs{\mathrm{d}}\tau \right)^2  +\frac{\bs{\mathrm{d}}r^2}{a(r)} 
    \\
    &\feq+ \tilde{d}(r)\left(\bs{\mathrm{d}}q +2nx\bs{\mathrm{d}}\tau\right)\vee\bs{\mathrm{d}}r\;,
\end{aligned}
\end{equation}
while the $\Gamma$-invariant field strength is given by
\begin{equation}\label{eq:F4310from435}
\begin{aligned}
    \bs{F}_{\text{[4,3,10]}} &= -\tilde{f}_1(r) \, \bs{\mathrm{d}}\tau \wedge \bs{\mathrm{d}}x \\
    &\quad + \tilde{f}_2(r) \left( \bs{\mathrm{d}}q + 2n x \bs{\mathrm{d}}\tau \right) \wedge \bs{\mathrm{d}}r\;.
\end{aligned}
\end{equation}

Introducing ${r=\hat{\rho}^2/(4n)}$, ${q=2n\hat{\varphi}}$, ${x=\hat{z}/n}$, ${\tau=\hat{t}/n}$, the metric \eqref{eq:4310from435} will take a more familiar form,
\begin{equation}\label{eq:4310alternative}
\begin{aligned}
    \bs{g}_{\text{[4,3,10]}} &= 4\hat{j} c(\xfrac{\sqrt{\hat{j}}\hat{\rho}^2}{2}) (-\bs{\mathrm{d}}\hat{t}^2+\bs{\mathrm{d}}\hat{z}^2) + \tfrac{\hat{j}\hat{\rho}^2\bs{\mathrm{d}}\hat{\rho}^2}{a(\frac{\sqrt{\hat{j}}\hat{\rho}^2}{2})} 
    \\
    &\feq+ \tfrac{a(\frac{\sqrt{\hat{j}}\hat{\rho}^2}{2}) b(\frac{\sqrt{\hat{j}}\hat{\rho}^2}{2})}{\hat{j}}\left(\bs{\mathrm{d}}\hat{\varphi} +4\hat{j}\hat{z}\bs{\mathrm{d}}\hat{t} \right)^2  
    \\
    &\feq+ \tfrac{\hat{\rho}\tilde{d}(\frac{\sqrt{\hat{j}}\hat{\rho}^2}{2})}{2\sqrt{\hat{j}}}\left(\bs{\mathrm{d}}\hat{\varphi} +4\hat{j}\hat{z}\bs{\mathrm{d}}\hat{t}\right)\vee\bs{\mathrm{d}}\hat{\rho}\;,
\end{aligned}
\end{equation}
where we denoted ${\hat{j}=1/(4n^2)}$. Now, it becomes obvious that the swirling universe is contained in the metric ansatz $\bs{g}_{\text{[4,3,10]}}$ [(3.3) from \cite{Astorino:2022aam}].

\section{Application to GR and beyond}\label{sc:application}

Although the mappings above are independent of a theory, it is instructive to show their usefulness and put our results in the context of previous literature. There are two important points to mention regarding their application. 

The double Wick rotations of the $\Gamma$-invariant tensors described above contain terms with tilde, which either have to vanish or must be redefined to real ones. In the $\Gamma$-invariant metrics $\bs{g}$, the tilde only appears above the $d$ terms, which can always be gauge-fixed to zero by means of \eqref{eq:resdifgen}. Hence, the vacuum solutions of any metric theory of gravity (assuming standard analyticity of the theory and the solutions in these coordinates) are simply mapped to the new vacuum solutions (of the same theory) only through the above analytic continuation of coordinates. However, the situation is different for the case of the $\Gamma$-invariant field strengths $\bs{F}$. In order to preserve their reality through the above analytic continuations, one has to ensure real $\tilde{f}_1$ and $\tilde{f}_2$, for example, by extra analytic continuation of the electromagnetic charges, or by some theory changes (e.g., continuing the couplings or redefining the field).

As we saw, there exist infinitesimal group actions that actually map to two other infinitesimal group actions under two different double Wick rotations, cf. [4,3,8] and [4,3,9] mapping respectively to [4,3,\{3,1\}] and [4,3,\{4,2\}] in Fig.~\ref{fig:enter-label}. However, due to the sign difference in $c(r)$, a chosen $\Gamma$-invariant metric with [4,3,8] maps only to one of the two [4,3,\{3,1\}] with the standard Lorentzian signature, while to the other with the reversed-sign Lorentzian signature and should be discarded. This depends on whether there is $\mathrm{dS_2}$ or $\mathrm{AdS_2}$ structure in the metric of [4,3,8]. Conversely, metrics from [4,3,3] and [4,3,1] yield two distinct geometries, both possessing the symmetries [4,3,8]. Also, there is no map of a $\Gamma$-invariant metric between [4,3,3] and [4,3,1], even though the two infinitesimal group actions are related by double Wick rotations. Analogous situation occurs for [4,3,9].

\subsection{\texorpdfstring{Einstein--Maxwell--$\Lambda$}{}}

Concerning GR, the ansatz \eqref{Ansatz}, \eqref{Ansatz2}, with ${n\neq0}$, i.e., [4,3,\{4,2,5\}], contains the spherical, hyperbolic, or planar charged TNUT--(A)dS solutions and its NUT-like subcases (e.g., massless hyperbolic TNUT, planar TNUT--(A)dS, etc.). These are given by
\begin{equation}\label{eq:TNUTs}
\begin{aligned}
    a &=\frac{k (r^2 {-} n^2) - 2 m r - \frac{\Lambda}{3} \left(r^4 {+} 6 n^2 r^2 {-} 3 n^4\right) + q_{\textrm{e}}^2+q_{\textrm{m}}^2}{r^2 + n^2}\;,
    \\
    b&=1\;,
    \quad 
    c=r^2+n^2\,,
    \quad 
    d=0\,,
\end{aligned}
\end{equation}
and
\begin{equation}\label{eq:TNUTsF}
    f_1 = \frac{q_{\textrm{m}}(r^2 {-} n^2) + 2q_{\textrm{e}}nr}{r^2 + n^2} \;, \;\; f_2 = - \frac{q_{\textrm{e}}(r^2 {-} n^2) + 2q_{\textrm{m}}nr }{(r^2 + n^2)^2} \;,
\end{equation}
where ${m,q_{\textrm{e}},q_{\textrm{m}},\Lambda\in\mathbb{R}}$ are parameters corresponding to mass, electromagnetic charges, and cosmological constant. On the other hand, the ansatz \eqref{Ansatz}, \eqref{Ansatz2}, with ${n=0}$, i.e., [4,3,\{3,1,6\}], includes the spherical, hyperbolic, or planar RN--(A)dS and its subcases (Schw., AII-(A)dS, etc.); these correspond again to \eqref{eq:TNUTs}, \eqref{eq:TNUTsF}, but upon setting ${n=0}$. Notice that the case ${k=0}$ (for any ${n\in\mathbb{R}}$) admits the scaling freedom: ${t\to t/S}$, ${r\to Sr}$, ${\rho\to\rho/S}$, ${n\to Sn}$, ${m\to S^3 m}$, ${q_{\textrm{e}}\to S^2q_{\textrm{e}}}$, ${q_{\textrm{m}}\to S^2q_{\textrm{m}}}$, for ${S>0}$, which allows setting $m$ to an arbitrary non-zero value.

Let us now go through the double Wick rotations of these electrovacuum solutions to demonstrate the above points explicitly. In all charged cases, the electromagnetic charges must be analytically continued as $\tilde{q}_{\textrm{e/m}}=iq_{\textrm{e/m}}$, which automatically renders $\tilde{f}_1$ and $\tilde{f}_2$ real. Other than that, the expressions in \eqref{eq:TNUTs}, \eqref{eq:TNUTsF}, remain unchanged. We also assume ${a>0}$, which corresponds to the stationary domain in the original metric.

We begin with the solutions given by \eqref{Ansatz}, \eqref{Ansatz2}, with \eqref{eq:TNUTs}, \eqref{eq:TNUTsF}, where ${n=0}$. The spherical RN--(A)dS solution (${k=1}$) can be mapped to the charged BI--(A)dS \eqref{eq:metric438fromk433}, \eqref{eq:F438fromk433} [same functions \eqref{eq:TNUTs}, \eqref{eq:TNUTsF} but with $\tilde{q}_{\textrm{e/m}}=iq_{\textrm{e/m}}$], while the hyperbolic RN--(A)dS solution (${k=-1}$) to the charged BII--(A)dS \eqref{eq:metric438from431}, \eqref{eq:F438from431}.\footnote{The results match (132) in \cite{Podolsky:2018dpr} upon ${\rho=p=\sqrt{1+\hat{q}^2}}$.} Although both solutions belong to [4,3,8], mapping charged BI--(A)dS to [4,3,1] or charged BII--(A)dS to [4,3,3] would result in a metric with reversed-sign Lorentzian signature. The planar RN--(A)dS solution (${k=0}$) can be mapped to charged BIII--(A)dS \eqref{eq:4311from436}, \eqref{eq:F4311from436}, i.e., [4,3,11].\footnote{Again, it corresponds to (132) from \cite{Podolsky:2018dpr}.} Remark that the uncharged BIII--(A)dS metric is a special cases of Levi--Civita ($\Lambda=0$) and Linet--Tian ($\Lambda\neq0$) metrics with ${\sigma=1/4}$ while the charged BIII--(A)dS is the Melvin--(A)dS spacetime.\footnote{The former is shown in \cite{Podolsky:2018dpr} while the later is seen by setting ${m=-(\tilde{q}_{\textrm{e}}^2+\tilde{q}_{\textrm{m}}^2)/2}$ using the scaling freedom and choosing ${\hat{B}=\sqrt{\tilde{q}_{\textrm{e}}^2+\tilde{q}_{\textrm{m}}^2}}$ in \eqref{eq:4311alternative}, which then matches (3.62) in \cite{Barrientos:2024pkt}.}

Moving on to the solutions given by \eqref{Ansatz}, \eqref{Ansatz2}, with \eqref{eq:TNUTs}, \eqref{eq:TNUTsF} where ${n\neq0}$. The massless charged hyperbolic TNUT--(A)dS (${k=-1}$, ${m=0}$) maps to NHEK--N--(A)dS \eqref{eq:439from432}, \eqref{eq:F439from432}, where the extra `N' stands for `Newman'.\footnote{By changing the coordinates ${\hat{r}= {2 r_+'^2r'}/{[(1+\xfrac{\Lambda}{3} a'^2  ) (a'^2+r_+'^2)]} }$, ${\cos\hat{\theta}= \hat{r}_0{(1+\frac{\Lambda}{3}a'^2 ) (a'^2+r_+'^2) \sqrt{1-\xfrac{\Lambda}{3}  (a'^2+6 r_+'^2)}}\sigma'/(2 \sqrt{2} r_+'^3)}$, ${\hat{\phi}{=}\xfrac{r_+'^2 \left(2 r_+'^2{-}q'^2{-}4 \Lambda  r_+'^4\right){-}\frac{\Lambda}{3}a'^4   q'^2{-}a'^2 \left(q'^2 \left(1{+}\frac{\Lambda}{3}  r_+'^2\right){+} \frac{2\Lambda}{3}  r_+'^4\right)}{2 r_+'^3\sqrt{ \left(1-\frac{\Lambda}{3}  \left(a'^2+6 r_+'^2\right)\right) \left(r_+'^2 \left(1-\frac{\Lambda}{3}  \left(a'^2+6 r_+'^2\right)\right)- q'^2\right)}} \phi'}$\!, and by rescaling the constant parameters  ${\hat{r}_0={\sqrt{2}r'_+}/{\sqrt{1{-}\xfrac{\Lambda}{3}  (a'^2{+}6 r_+'^2)}} }$, ${\tilde{q}_{\textrm{e/i}}={q'_{\textrm{e/i}}}/{[1{-}\xfrac{\Lambda}{3}  (a'^2{+}6 r_+'^2)]}}$, one can recast \eqref{eq:439alt} to the match exactly (54) with (61)--(66) from \cite{Kunduri:2008tk}, where we introduced ${q'^2=q_{\textrm{e}}'^2{+}q_{\textrm{i}}'^2}$, ${a'^2{=}{r_+'^2-q'^2-\Lambda  r_+'^4}/{(1+\xfrac{\Lambda}{3}  r_+'^2)}}$.}
On the other hand, the massless charged spherical TNUT--(A)dS (${k=1}$, ${m=0}$) maps to a spacetime \eqref{eq:439from434}, \eqref{eq:F439from434}, which do not appear to be discussed in the literature. As mentioned above, we expect it to correspond to the near-horizon extreme limit of a rotating black hole with hyperbolic horizon topology \cite{Klemm:1997ea}. As before, although both solutions belong to [4,3,9], mapping them in opposite would lead to a metric with reversed-sign Lorentzian signature. Remark that a different analytic continuation exists in the literature \cite{Kunduri:2008tk}, which maps the spherical TNUT to the standard (spherical topology) NHEK instead, but it requires analytic continuation of other coordinates and parameters. The charged planar TNUT--(A)dS (${k=0}$) maps to swirling--Melvin--(A)dS \eqref{eq:4310from435}, \eqref{eq:F4310from435}, i.e., [4,3,10].\footnote{This can be seen by setting ${m=-\sqrt{B'^4+16j'^2}/(16j'^{3/2})}$ using the scaling freedom, where we introduced ${B'=4j'\sqrt{\tilde{q}_\textrm{e}^2+\tilde{q}_\textrm{m}^2}}$, ${\hat{j}=B'^4/(16j')+j'}$. Then, \eqref{eq:4310alternative} can be transformed using ${\hat{\rho}=\sqrt{4B'/(B'^4+16j'^2) +\rho'^2}}$, ${\hat{t}=4j't'/\sqrt{B'^4+16j'^2}}$, ${\hat{z}=4j'z'/\sqrt{B'^4+16j'^2}}$ to a form matching (3.63) in \cite{Barrientos:2024pkt}.}

\subsection{\texorpdfstring{Einstein--ModMax--$\Lambda$}{}}

To show that the above double Wick rotation extends beyond GR, we present an example from NLE. The ansatz \eqref{Ansatz}, \eqref{Ansatz2}, with ${n\neq0}$, i.e., [4,3,\{4,2,5\}], also includes the charged spherical/hyperbolic/planar TNUT-(A)dS-type solution in Einstein--ModMax--$\Lambda$ \cite{Flores-Alfonso:2020nnd,BallonBordo:2020jtw}, where now
\begin{equation}\label{eq:TNUTModMax}
\begin{aligned}
    a &=\big[k (r^2 {-} n^2) - 2 m r - \tfrac{\Lambda}{3} \left(r^4 {+} 6 n^2 r^2 {-} 3 n^4\right) \\
    &\feq+ e^{-\gamma}(q_{\textrm{e}}^2+q_{\textrm{m}}^2)\big]/(r^2 + n^2)\;,
    \\
    b&=1\;,
    \quad 
    c=r^2+n^2\,,
    \quad 
    d=0\,,
\end{aligned}
\end{equation}
and
\begin{equation}\label{eq:TNUTModMaxF}
\begin{aligned}
    f_1 &= -q_{\textrm{e}} \sin (e^{-\gamma}(\pi - 2 \arctan \tfrac{r}{n}))\\ &- q_{\textrm{m}} \cos (e^{-\gamma}(\pi - 2 \arctan \tfrac{r}{n}))\;, \quad f_2 = - \tfrac{1}{2n} f'_1\;.
\end{aligned}
\end{equation}
The planar case ${k=0}$ (i.e., [4,3,5]) maps correctly to the swirling--Melvin--(A)dS-type solution (i.e., [4,3,10]) \eqref{eq:4310from435}, \eqref{eq:F4310from435} [same functions \eqref{eq:TNUTModMax}, \eqref{eq:TNUTModMaxF} but with $\tilde{q}_{\textrm{e/m}}=iq_{\textrm{e/m}}$].\footnote{This can be seen by performing the coordinate transformation $r= -S\rho'^2/(4n) - n\sqrt{S-1}$, $q = 2nS^{-1} \varphi'$, $x = S^{-\frac12} z'/n$, $\tau = S^{-\frac12} t'/n$ and setting $m = \frac{n}{2}S^2$, $q_{\textrm{e}}^2 + q_{\textrm{m}}^2 = -n^4 S^2 (E'^2+B'^2)$, $S = 1 + e^{-2\gamma}n^4(E'^2+B'^2)^2$, $n = 1/(2\sqrt{j'})$ to recover (54), (55) in \cite{Barrientos:2024umq}.}

\section{Conclusions}\label{sc:concl}
In the present work, we explored mutual relations of distinct infinitesimal group actions $\Gamma$ corresponding to certain 4-dimensional Lie algebras of Killing vectors with 3-dimensional orbits as summarized in Fig.~\ref{fig:enter-label}. Specifically, we showed that symmetries of spherical, hyperbolic, and planar TNUT spacetimes map under double Wick rotations of coordinates to symmetries of NHEK or swirling spacetimes. Similarly, the static spherical, hyperbolic, and planar symmetries, i.e., the symmetries of A-metrics, map to symmetries of B-metrics or Melvin. This was naturally translated to relations within the corresponding classes of $\Gamma$-invariant metrics and classes of $\Gamma$-invariant electromagnetic field strengths. All the resulting relations are theory independent. Curiously, all the double-Wick-related infinitesimal group actions [4,3,\{1--6,8--11\}] (and only those among all [4,3,1--20]) satisfy the principle of symmetric criticality \cite{Frausto:2024egp}, i.e., they allow consistent symmetry reduction of Lagrangians.

We intentionally avoided any transformations of the coordinate $r$ labeling the orbits (except for comparison with the literature) and only performed the complex analytic continuations of coordinates within the orbits. Also, analytic continuations of the parameter $n$ were unnecessary, which is somewhat artificial at the level of infinitesimal group actions: only ${n = 0}$ versus ${n \neq 0}$ is relevant. Regarding the vacuum solutions in an arbitrary metric theory of gravity, there is no need for analytic continuation of any other coordinates or parameters. As a result, the expressions for functions $a(r)$, $b(r)$, and $c(r)$ (in the gauge ${d(r)=0}$) remain completely unchanged while the vacuum solutions map to vacuum solutions with different symmetries. The above can also be extended to theories with an electromagnetic field, provided the double Wick rotations are supplemented by other modifications --- e.g., analytic continuation of the charges --- to ensure $\tilde{f}_1$ and $\tilde{f}_2$ remain real; this then affects $a(r)$, $b(r)$, and $c(r)$ as well.

We commented on the relationship of specific electrovacuum solutions in Einstein--Maxwell--$\Lambda$ theory (see labels in Fig.~\ref{fig:enter-label}) as well as in Einstein--ModMax--$\Lambda$ theory. The former has uncovered some solutions that may not have been explicitly written in the literature, such as the potential near-horizon extreme limits of rotating black holes with hyperbolic horizon topology. Applying all these mappings to known solutions (e.g., known TNUT-type solutions) of modified theories of gravity should generate new solutions exhibiting the symmetries discussed above. It would also be interesting to apply these relations in other theories, e.g., in Einstein--Born--Infeld to rederive the NHEK solution from \cite{Hale:2025urg}.

Another worthwhile direction would be to investigate other possible relations within the infinitesimal group actions in the Hicks classification. In fact, even the cases we have studied [4,3,\{1--6,8--11\}] can also be viewed from different viewpoints. For example, [6,4,\{1--5\}] are symmetries of direct product spacetimes such as the Bertotti--Robinson, Nariai, etc., and are already included in various ways as special cases of [4,3,\{1,3,6,8,11\}] with extra symmetries (subclasses of $\Gamma$-invariant metrics) \cite{Frausto:2024egp}. On the other hand, the symmetries [4,3,\{2,4,5,9,10,11\}] correspond to the special cases of [3,3,\{2,3,8,9\}] describing the symmetries of Bianchi I, II, VIII, and IX cosmologies.

\begin{acknowledgments}
A.C. and I.K. acknowledge financial support from the Primus grant PRIMUS/23/SCI/005 of Charles University. I.K. further thanks the Charles University Research Center grant UNCE24/SCI/016 for support. T.M. is supported by the Czech Science Foundation (GAČR) grant No.~25-15544S.
\end{acknowledgments}

\appendix

\section{Hicks classification}\label{sc:Hicks}

Let $M$ be a 4-dimensional manifold, and let $\Gamma$ be a $d$-dimensional Lie algebra of vector fields on $M$, generating the infinitesimal action of a Lie group $G$. Note that the Lie algebra of vector fields $\Gamma$ carries more information than its abstract counterpart $\mathcal{G}$ defined solely by the structure constants (independent of $M$). A given $\mathcal{G}$ (generating $G$) may act in various ways on the same manifold $M$ and lead to its different realizations $\Gamma$. Individual infinitesimal group actions $\Gamma$ can then be distinguished by analyzing their isotropy subalgebras  ${\mathcal{G}_{\mathrm{x}}\subseteq\mathcal{G}}$ realized by the vector fields ${\Gamma_{\mathrm{x}} = \{\,\bs{X} \in \Gamma \mid \bs{X}|_{\mathrm{x}} = 0\,\} \subseteq \Gamma}$ (i.e., the infinitesimal action of the isotropy subgroup ${G_\mathrm{x} = \{ \gamma \in G \mid \gamma \mathrm{x} = \mathrm{x} \}\subseteq G}$), which consists of those that leave the point ${\mathrm{x} \in M}$ unchanged. Remark that ${\Gamma/\Gamma_{\mathrm{x}}}$ is the tangent space (at $\mathrm{x}$) to the orbit of $\mathrm{x}$ (i.e., $\{ \gamma \mathrm{x}\in M \mid \gamma \in G \}\subseteq M$, which is diffeomorphic to the homogeneous space $G/G_{\mathrm{x}}$). 

If $\Gamma$ preserves a Lorentzian metric $\bs{g}$, ${\lie_{\bs{X}}\bs{g}=0}$ for all ${\bs{X}\in \Gamma}$,  then $\Gamma$ is the Lie algebra of Killing vectors (the generators of isometries) and $\bs{g}$ is called the $\Gamma$-invariant metric, but one may construct $\Gamma$-invariant tensors of any type the same way. The isotropy subalgebra $\mathcal{G}_{\mathrm{x}}$ acts on $\mathcal{G}/\mathcal{G}_{\mathrm{x}}$ via the induced adjoint representation, which induces an action on the tangent space to the orbit ${\Gamma/\Gamma_{\mathrm{x}}}$. Since $\Gamma$ are the Killing vectors of the Lorentzian metric $\bs{g}$, the isotropy subalgebras $\mathcal{G}_{\mathrm{x}}$ correspond to the subalgebras of the Lorentz algebra.

Distinct infinitesimal group actions $\Gamma$ can be classified by the Lorentzian Lie algebra-subalgebra pairs $(\mathcal{G},\mathcal{G}_{\mathrm{x}})$ with $\mathcal{G}_{\mathrm{x}}$ being a distinguished subalgebra of $\mathcal{G}$ acting on ${\mathcal{G}/\mathcal{G}_{\mathrm{x}}}$ as the subalgebra of Lorentz algebra.\footnote{Two pairs are equivalent if there exists a Lie algebra isomorphism ${\phi:\mathcal{G}\to\mathcal{G}'}$ with ${\phi(\mathcal{G}_{\mathrm{x}})=\mathcal{G}'_{\mathrm{x}}}$.} This holds whenever the manifold locally splits into an isotropy-invariant slice and a group orbit, with isotropy subalgebras conjugate under the adjoint action throughout the neighborhood (e.g., the isometry groups with generically non-null orbits). We refer to the classification by abstract Lorentzian pairs as the Hicks classification \cite{Hicks:thesis} (see also \cite{Frausto:2024egp}), which completes previous classifications along these lines \cite{Fels_Renner_2006,Bowers:2012,Snobl2014-te,Rozum2015-lp} and significantly improves upon the original Petrov classification of Killing vector fields \cite{Petrov}.\footnote{This should not be confused with the well-known Petrov classification based on the algebraic properties of the curvature.} The individual infinitesimal group actions $\Gamma$ in this classification are denoted by the triplet ${[d,l,c]}$, which respectively captures the dimension of $\mathcal{G}$, the dimension of the orbit ${l=d-p}$ where $p$ is the dimension of $\mathcal{G}_{\mathrm{x}}$, and an additional distinguishing label $c$. In this paper, we are primarily interested in certain cases within [4,3,-] that represent a very interesting symmetries from the viewpoints of both physical applications and mathematical tractability.


%

\end{document}